\documentclass[12pt,oneside]{article}
\usepackage{amsfonts,amssymb,graphicx}
\def\be{\begin{equation}}
\def\ee{\end{equation}}
\def\bea{\begin{eqnarray}}
\def\eea{\end{eqnarray}}
\def\<{\langle}
\def\>{\rangle}
\def\~{\tilde}
\def\s{\sigma}

\def\a{\alpha}
\def\b{\beta}

\def\o{\omega}
\def\t{\tau}

\newcommand{\B}{\Bbb B}

\newcommand{\Z}{\Bbb Z}

\newcommand{\av}[1]{\mbox{{\rm Av}}\left(#1\right)}

\newtheorem{theorem}{Theorem}
\newtheorem{lemma}{Lemma}

\begin{document}
\begin{center}
{\bf\sc\Large
on the surface pressure for \\the edwards-anderson model}\\
\vspace{1cm}
{Pierluigi Contucci, Sandro Graffi}\\
\vspace{.5cm}
{\small Dipartimento di Matematica} \\
    {\small Universit\`a di Bologna,
    40127 Bologna, Italy}\\
    {\small {e-mail: contucci@dm.unibo.it, graffi@dm.unibo.it}}\\
\vskip 1truecm
{\small June 3th, 2003. Revised: February 11th, 2004}
\end{center}
\vskip 1truecm
\begin{abstract}\noindent
For the Edwards-Anderson model we introduce an integral representation for
the 
{\it surface pressure} (per unit surface) $\t_{\partial\Lambda}$ in terms of
a quenched 
moment of the bond-overlap on the surface. We consider free $\Phi$, periodic
$\Pi$ and
antiperiodic $\Pi^*$ boundary conditions (by symmetry
$\t^{(\Pi)}_{\partial\Lambda}=\t^{(\Pi^*)}_{\partial\Lambda}$),
and prove the bounds
\bea\nonumber
-\frac{1}{4} \; \le \; \t^{(\Phi)}_{\partial\Lambda} \; \le 0 \; ,
\eea
\bea\nonumber
\t^{(\Phi)}_{\partial\Lambda}\; \le \; \t^{(\Pi)}_{\partial\Lambda} \; \le
\frac{1}{2} \; ,
\eea
We show moreover that at high temperatures $\t^{(\Phi)}_{\partial\Lambda}$ is
close to $-\beta^2/4$ and
$\t^{(\Pi)}_{\partial\Lambda}$ is close to $\beta^2/4$ uniformly in the volume
$\Lambda$. 
\end{abstract}
\newpage\noindent
\section{Introduction}
In statistical mechanics once the existence of the thermodynamic limit has
been proved
for the free energy per unit volume a natural subsequent
question is to establish at which rate with respect to the volume such
limit is reached. In particular it is interesting to determine  the next
term in the expansion
\bea\nonumber
\ln Z_\Lambda \; = \; p|\Lambda| + o(|\Lambda|) \; .
\eea
The problem has been analyzed since the pioneering work by Fisher and
Lebowitz \cite{FL}
on classical particle systems and followed by a series of results in both
Euclidean quantum field theories \cite{G,GRS} and in ferromagnetic spin
models \cite{FC}. 
In those cases  the basic properties
of monotonicity and convexity of the thermodynamic quantities with respect
to the strength 
of the interaction, namely the first and second Griffiths inequalities, 
made possible a rigorous
proof of what thermodynamics suggests (see \cite{Si}): for sufficiently
regular potentials and
(say) free boundary conditions the pressure varies with the volume as
\be
\ln Z_\Lambda\, = \, p|\Lambda| + \tau|\partial\Lambda|
+o(|\partial\Lambda|) \; ,
\label{th}
\ee
where $p$ is the thermodynamic limit of the pressure per unit volume
\be
p_\Lambda \, = \, \frac{\ln Z_\Lambda}{|\Lambda|} \; ,
\label{pr}
\ee
and $\t$ is the thermodynamic limit of the {\it surface pressure} per unit
area
\be
{\tau}_{\partial\Lambda} \, = \, \frac{\ln Z_\Lambda -
p|\Lambda|}{|\partial\Lambda|} \; .
\label{sp} 
\ee
The quantity $\t$, unlike $p$, depends in general not only on the  interaction 
but also on the boundary conditions and represents
the contribution to the pressure due to the interaction of the system with
its boundary.

In this paper we analyze the surface pressure problem for the
Edwards-Anderson model
with Gaussian couplings in the quenched ensemble. Basing on the property of
existence,
self averaging and independence on the boun\-dary conditions of the
thermodynamic limit
for the random pressure per particle (see for instance
\cite{VV} and \cite{CG}) we study the correction to the leading term for
different
boundary conditions (free, periodic and antiperiodic). Our main idea relies
on an inequality 
which translates to random systems the contents of the {\it first} Griffiths
inequality: in a ferromagnet the free
energy decreases with the strength of each interaction, in a spin-glass the
free energy decreases with the
variance of each random coupling. Our technical tool
is an interpolation method (similar to those in \cite{GT} and \cite{CG})
which plays, in spin glass statistical mechanics, the same
role of the Griffiths interpolation method \cite{Si,Gr}  in classical
ferromagnetic systems. Our main result is an integral
representation theorem for the surface pressure in the quenched ensemble
for different boundary conditions and  rectangular boxes.
As an immediate consequence we find that its value is bounded from above by
$0$ and from below by $-1/4$ and that for high temperatures it is non-zero. We
prove moreover that the surface pressure for periodic or antiperiodic boundary
conditions is larger than the free one and we provide
an integral representation for their quenched difference which we control at
high temperature.

\section{Definitions and Results}
Consider the Edwards-Anderson $d$-dimensional spin-glass model defined by
configurations
of Ising spins $\s_n$, $n\in \Lambda\subset \Z^d$ for some $d$-parallelepiped
$\Lambda$. To be definite  we locate it in the positive quadrant of $\Z^d$ with 
a vertex in the origin. We
denote $L_1,L_2,...,L_d$ the sides, $|\Lambda|$ the volume and   $|\partial 
\Lambda|$ the surface. The 
interaction is described by the potential
\be
U_{\Lambda}(J,\s) \, = \, \sum_{(n,n')\in B(\Lambda)}J_{n,n'}\s_n\s_{n'} \;
,
\label{eag}
\ee
where the $J_{n,n'}$ are independent normal Gaussian variables and the sum
runs over all  pairs of nearest neighbors sites $|n-n'|=1$. We use here
the standard identification of
the space of nearest neighbors with the $d$-dimensional {\it bond}-lattice
$b\in \B^d$ with $b=(n,n')$
and denote $B(\Lambda)$ the $d$-bond-parallelepiped associated to $\Lambda$.
Given two spin configurations   $\s$ and $\t$ 
introduce the notation
$\s_b=\s_n\s_{n'}$ and $\t_b=\t_n\t_{n'}$;  the local bond-overlap between
$\s$ and $\t$  is
\be
q_b(\s,\t):=\s_b\t_b \; ;
\ee
for every $B\subset B(\Lambda)$ we define
\be
q_B(\s,\t) := \frac{1}{|B|}\sum_{b\in B} q_b(\s,\t)\; .
\label{bov}
\ee
The reason to introduce the bond overlap is related to the
mathematical structure of the Hamiltonian (\ref{eag}): as a sum of
Gaussian variables it
is, for each $\s$-configuration, a Gaussian variable itself and thus by the 
Wick theorem completely identified by
its covariance matrix which is proportional to the bond-overlap
$q_{B(\Lambda)}(\s,\t)$.
Denoting $Av$ the Gaussian average we have in fact:
\bea\nonumber
Av(U_{\Lambda}(J,\s) U_{\Lambda}(J,\t)) &=& \sum_{b,
b'}Av(J_{b}J_{b'})\s_{b}\t_{b'}  \\
&=& \sum_{b, b'} \delta_{b,b'}\s_{b}\t_{b'} = |B(\Lambda)| 
q_{B(\Lambda)}(\s,\t)
\; .
\label{cova}
\eea
{\bf Definitions}. \\
For assigned boundary conditions $\Xi$ we consider
\begin{enumerate}
\item the random partition function,
\be
Z^{(\Xi)}_\Lambda(J) \, := \, \sum_{\s} e^{U^{(\Xi)}_\Lambda(\s,J)}
\; ,
\ee
\item the random pressure
\be
P^{(\Xi)}_\Lambda(J) \, := \, \ln Z^{(\Xi)}_\Lambda(J) \; ,
\ee
\item the quenched pressure
\be 
P^{(\Xi)}_\Lambda := \, \av{\ln Z^{(\Xi)}_\Lambda(J)} \; ,
\label{fe}
\ee
\item
the product (over the same disorder) random Gibbs-Boltzmann state
\be
\omega^{(\Xi)}_{\Lambda}(-) \, := \,
\frac{\sum_{\sigma,\t}-\,e^{U^{(\Xi)}_\Lambda(\s)+U^{(\Xi)}_\Lambda(\t)}}
{[Z^{(\Xi)}_\Lambda]^2}
\; ,
\label{omega}
\ee
\item the quenched equilibrium state
\be
<->^{(\Xi)}_\Lambda \, := \av{\omega(-)^{(\Xi)}_\Lambda} \; ,
\ee
\item the random surface pressure
\be
T^{(\Xi)}_{\Lambda}(J) \, := \, P^{(\Xi)}_\Lambda(J) - p|\Lambda| \; ,
\label{qsp} 
\ee
\item and the quenched surface pressure
\be
T^{(\Xi)}_{\Lambda} \, := \,  \av{T^{(\Xi)}_{\Lambda}(J)} \; .
\ee
\end{enumerate}
We will consider three types of boundary conditions. The free ones $\Phi$ in
which
the partition sum runs over all the spins inside the parallelepiped
$\Lambda$:
\be
Z^{(\Phi)}_\Lambda(J) \, := \, \sum_{\s} e^{U_{\Lambda}(\s,J)} \; .
\ee
The periodic boundary conditions $\Pi$ in which the partition sum runs over
all the spin values in the
torus $\Pi_\Lambda=\Z^d/\Lambda$:
\be
Z^{(\Pi)}_\Lambda(J) \, := \, \sum_{\s} e^{U_{\Pi_\Lambda}(\s,J)} \; .
\ee
The anti-periodic conditions $\Pi^*$ are defined summing over the spin 
configurations
with the condition, for instance in $d=1$, that $\s_1=-\s_{N+1}$. This is 
clearly
equivalent, for a given choice of $J$, to consider a sistem with periodic 
boundary 
conditions and with a changed sign of $J_{1,N+1}$. In $d$ dimensions the 
general definition
is given as follows: consider the standard orthogonal cut of the torus which 
unfolds $\Pi$ to
$\Lambda$ i.e. the set $\partial
B(\Lambda)$ defined as the
collection of $b=(n,n')$ with $n<n'$ (according to the lexicographic order) and 
$n=(n_1,n_2,...,n_k)$ in which
$n_i=1$ $\forall i\neq k$ and $n_k=0$. 
Given
\be
\a_b \; = \; \left\{
\begin{array}{ll} 
-1, & \mbox{if $b\in\partial B(\Lambda)$}, \\
 1, & \mbox{otherwise} \, ,
\end{array}\right. 
\ee
and the potential
\be
U_{\Pi^{*}_\Lambda}(\s,J) \, = \, \sum_{b\in B(\Pi_\Lambda)}\a_bJ_b\s_b \; ,
\ee
the anti-periodic boundary condition partition sum runs over all the spins in 
the torus
$\Pi_\Lambda=\Z^d/\Lambda$
\be
Z^{(\Pi^*)}_\Lambda(J) \, := \, \sum_{\s} e^{U_{\Pi^{*}_\Lambda}(\s,J)} \; .
\ee
To state our results we first establish some further notation.
Consider the boundary bond-overlap
\be
q_{\partial B(\Lambda)} \; = \; \frac{1}{|\partial
B(\Lambda)|}\sum_{b\in\partial  B(\Lambda)} q_b \; .
\ee
Let $k\Lambda$ be the k-magnified $\Lambda$ defined, for each
positive integer $k$, as the d-parallelepiped of
sides $kL_1,kL_2,...,kL_d$ and consider the magnificated torus 
\be
\Pi_{k\Lambda}=\Z^d/k\Lambda \; .
\ee
Define the set
\be
{\cal C}_{\Pi_{k\Lambda}} \; := \; B(\Pi_{k\Lambda}) \backslash
\bigcup_{s=1}^{k^d}B(\Lambda_s) \; .
\ee
and associate with $\Pi_{k\Lambda}$ the interpolating potential
\be
U_{\Pi_{k\Lambda}}(t) \, = \, \sum_{b\in B(\Pi_{k\Lambda})}
\sqrt{t_b}J_b\s_b \; ,
\label{iipp}
\ee
with
\be
t_b \; = \; \left\{
\begin{array}{ll} 
t, & \mbox{if $b\in {\cal C}_{\Pi_{k\Lambda}}$}, \\
1, & \mbox{otherwise},
\end{array}\right.
\ee
Finally let $<->^{(\Pi_{k\Lambda})}_t$ be the corresponding quenched state.

\begin{theorem}[Integral representation for
$T^{(\Phi)}_{\Lambda}$]\label{main}
The surface pressure per unit surface admits the representation
\be
T^{(\Phi)}_{\Lambda}  \, = \, - \frac{|\partial\Lambda|}{4}
\lim_{k\to\infty}\int_{0}^{1}
\left(1 - <q_{\partial B(\Lambda)}>^{(\Pi_{k\Lambda})}_t\right)dt \; ;
\label{azz1}
\ee
in particular the quantity
\be
\t^{(\Phi)}_{\partial\Lambda} \, = \,
\frac{T^{(\Phi)}_{\Lambda}}{|\partial\Lambda|}
\ee
admits the bounds
\be
-\frac{1}{4} \; \le \; \t^{(\Phi)}_{\partial\Lambda} \; \le \; 0 \; .
\label{opi}
\ee
\end{theorem}
\begin{theorem}[Integral representation for $T^{(\Pi)}_{\Lambda}$ and
$T^{(\Pi^*)}_{\Lambda}$]\label{main2}
For every $\Lambda$ the symmetry of the Gaussian distribution implies
\be
T^{(\Pi)}_{\Lambda} \, = \, T^{(\Pi^*)}_{\Lambda} \; .
\label{pippa}
\ee
Consider in the torus $\Pi_\Lambda$ the interpolating potential
\be
U^{(\Pi_{\Lambda})}(t) \, = \, \sum_{b\in B(\Pi_{\Lambda})}
\sqrt{t_b}J_b\s_b \, ,
\label{aip}
\ee
with
\be
t_b \; = \; \left\{
\begin{array}{ll} 
t, & \mbox{if $b\in\partial B(\Lambda)$}, \\
 1, & \mbox{otherwise} \, ,
\end{array}\right. 
\ee 
and let $<->^{(\Pi_\Lambda)}_t$ be its quenched state . Then
\be
T^{(\Pi)}_{\Lambda} \, = \, T^{(\Phi)}_{\Lambda} \, + \,
\frac{|\partial\Lambda|}{2}\int_{0}^{1}(1-<q_{\partial
B(\Lambda)}>^{(\Pi_\Lambda)}_t)dt \; .
\label{azz1}
\ee
In particular the quantity
\be
\t^{(\Pi)}_{\partial\Lambda} \, = \,
\frac{T^{(\Pi)}_{\Lambda}}{|\partial\Lambda|}
\ee
admits the bounds
\be
\t^{(\Phi)}_{\partial\Lambda} \; \le \; \t^{(\Pi)}_{\partial\Lambda} \; \le
\; \frac{1}{2}\; .
\label{opi}
\ee
\end{theorem}
\begin{theorem}[High temperatures]\label{ht}
Consider the potential
\be\label{tempe}
U_{\Lambda}(J,\s) = \beta\sum_{(n,n')\in B(\Lambda)}J_{n,n'}\s_n\s_{n'}
\ee 
Then:
\begin{itemize}
\item[{\rm (1)}]
There exist 
$\overline\b$ and $C>0$ depending only on $d$ such that for all $\b\le
\overline\b$
\be
\frac{\t^{(\Phi)}_{\partial\Lambda}}{\beta^2} \; \leq \; -C \; <\;0
\label{m1}
\ee
\item[{\rm (2)}]
For any $\varepsilon >0$ there exists $\b^{(\varepsilon)}>0$ such that
for all $\b\le \b^{(\varepsilon)}$
\be
\frac{\t^{(\Phi)}_{\partial\Lambda}}{\beta^2} \; \le \; 
-\frac{1}{4}(1-\varepsilon) \; ,
\label{m1}
\ee
and equivalently 
\be
\frac{\t^{(\Pi)}_{\partial\Lambda}}{\beta^2} \; \ge \; 
\frac{1}{4}(1-\varepsilon) \; ,
\label{m2}
\ee
uniformly in $\Lambda$.
\end{itemize}
\end{theorem}
\section{ Proof of the results}
We start by stating and proving the basic result.
\begin{lemma}[Monotonicity in the variance]\label{gsg} Let $t_b\ge 0$
$\forall b\in B(\Lambda)$ and $J_b$ be a normal Gaussian variable.
The Gaussian variable $\sqrt{t_b}J_b$ has variance $t_b$. Consider the
potential
$U_\Lambda=\sum_{b\in  B(\Lambda)}\sqrt{t_b}J_b\s_b$ with its associated
quenched thermodynamics. The quenched pressure $P_\Lambda$
is monotone increasing with respect to all the variances $t_b$:
\be
\frac{d}{d t_b}P_{\Lambda} \; = \; \frac{1}{2\sqrt{t_b}}\av{J_b\o(\s_b)} \;
= \; \frac{1}{2}(1 -<q_b>) \; \ge \; 0 \; .
\label{sega}
\ee
\end{lemma}
{\bf Proof of Lemma \ref{gsg}}.\\
The first equality is the chain rule on the
logarithm of an exponential of a square root:
\be
\frac{d}{d t_b}P_{\Lambda} \; = \; \frac{1}{2\sqrt{t_b}}\av{J_b
\frac{\sum_{\s}\s_be^{U(\s)}}{\sum_{\s}e^{U(\s)}}}
\label{mv}
\ee
Next we recall the integration by parts formula for normal Gaussian variables
\be\label{ibp}
\av{Jf(J)} \, = \, \av{\frac{df(J)}{dJ}} \; ,
\ee
the correlation derivative formula
\be\label{cd}
\frac{d\omega(\s_b)}{dJ_b} \, = \, \sqrt{t_b}\left(1 - \omega(\s_b)^2\right)
\;
\ge 0 \; ,
\ee
and the identity
\be
\omega(\s_b)^2 \, = \, \left(\frac{\sum_\s \s_b e^{U(\s)}}{\sum_\s
e^{U(\s)}}\right)^2 \, = \,
\frac{\sum_{\s,\t} \s_b\t_b e^{U(\s)+U(\t)}}{\sum_{\s,\t}
e^{U(\s)+U(\t)}} \, = \, \o(q_b) \; .
\label{cpi}
\ee
By applying successively (\ref{ibp}), (\ref{cd}) and (\ref{cpi}) we obtain
lemma \ref{gsg}.
\\ \\
{\bf Proof of Theorem \ref{main}}.\\
Given the $d$-parallelepiped $\Lambda$ consider  its
magnification $k\Lambda$ defined, for each positive integer $k$, as the
$d$-parallelepiped of
sides $kL_1,kL_2,...,kL_d$. Clearly $k\Lambda$ and $\Pi_{k\Lambda}$ are
partitioned
into $k^d$  non-empty disjoint cubes
$\Lambda_s$ all congruent to $\Lambda$ as explained in the definitions before 
Theorem
\ref{main}.
In finite volume and with free boundary
conditions we have by definition
\be
P_\Lambda^{(\Phi)} \; = \; \av{\ln Z_\Lambda} \; = \; k^{-d}\av{\ln
Z_\Lambda^{k^d}}\; .
\label{rep}
\ee
The limiting pressure per particle is independent on the
boundary conditions. Hence: 
\be
p|\Lambda| \; = \; \lim_{k\to\infty} k^{-d} \av{\ln Z^{(\Pi)}_{k\Lambda}} \;
\label{per}
\ee
By (\ref{rep}) and (\ref{per}) we obtain
\bea\nonumber
T^{(\Phi)}_{\Lambda} \, &=&  \left( P^{(\Phi)}_\Lambda - p|\Lambda| \right) \\
&=& \, \lim_{k\to \infty} k^{-d}\av{\ln Z_\Lambda^{k^d}- \ln
Z^{(\Pi)}_{k\Lambda}} \; .
\label{bs}
\eea
For each $0\le t\le 1$ we define the interpolating potential as in (\ref{iipp})
with 
\be
t_b \; = \; \left\{
\begin{array}{ll} 
t, & \mbox{if $b\in {\cal C}_{\Pi_{k\Lambda}}$}, \\
1, & \mbox{otherwise},
\end{array}\right.
\ee
the interpolating partition function
\be
Z^{(\Pi_{k\Lambda})}(t) \, = \, \sum_{\s} e^{U^{(\Pi_{k\Lambda})}(t)} \; ,
\ee
the interpolating pressure
\be 
P^{(\Pi_{k\Lambda})}(t) := \, \av{\ln Z^{(\Pi_{k\Lambda})}(t)} \; ,
\label{fe}
\ee
and the corresponding states $\o^{(\Pi_{k\Lambda})}_t(-)$ and
$<->^{(\Pi_{k\Lambda})}_t$.
We observe that
\be
\label{ifa}
Z^{(\Pi_{k\Lambda})}(0) \, = \, \prod_{s=1}^{k^d}Z_{\Lambda_s} \; , \; \quad
Z^{(\Pi_{k\Lambda})}(1)=Z_{\Pi_{k\Lambda}}\; ,
\label{st}
\ee
or equivalently 
\be
\label{ifa}
P^{(\Pi_{k\Lambda})}(0)  \, = \, k^dP_{\Lambda} \; , \; \quad
P^{(\Pi_{k\Lambda})}(1) \, = \, P_{\Pi_{k\Lambda}} \; ,
\ee
and by (\ref{bs}) 
\be
T^{(\Phi)}_{\Lambda} \, = \, \lim_{k\to \infty}
k^{-d}\left[P^{(\Pi_{k\Lambda})}(0)-P^{(\Pi_{k\Lambda})}(1)\right] \, = \,
- \lim_{k\to \infty} k^{-d} \int_{0}^{1}\frac{d}{dt}P^{(\Pi_{k\Lambda})}(t)
dt \; .
\label{bs2}
\ee
We remark now that
\be\label{derif} 
\frac{d}{dt}P^{(\Pi_{k\Lambda})}(t)\,=\, \sum_{b\in{\cal
C}_{\Pi_{k\Lambda}}}
\frac{1}{2\sqrt{t}}<J_{b}\s_b>^{(\Pi_{k\Lambda})}_t
\; ,
\ee 
and by Lemma \ref{gsg}
\be
\frac{d}{dt}P^{(\Pi_{k\Lambda})}(t) \, = \, \frac{1}{2}\sum_{b\in{\cal
C}_{\Pi_{k\Lambda}}}(1-<q_b>^{(\Pi_{k\Lambda})}_t) \; .
\ee
The  translation symmetry over the torus and the equality
$$
2|{\cal C}_{\Pi_{k\Lambda}}|=k^d|B(\partial\Lambda)|
$$ 
imply by
(\ref{bs2})
\be
\tau^{(\Phi)}_{\partial\Lambda} = -  \frac{1}{4} \lim_{k\to
\infty}\int_{0}^{1}
\left(1 - <q_{\partial B(\Lambda)}>_t^{(\Pi_{k\Lambda})}\right)dt \; .
\ee
\\ \\
{\bf Proof of Theorem \ref{main2}}. We first notice that the potential
\be
U^{(\Pi_{\Lambda})}(\a,\s,J) \, = \, \sum_{b\in B(\Pi_{\Lambda})}
\a_bJ_b\s_b \, ,
\ee
has a quenched pressure independent of $\a$ for each choice of $\a_b=\pm 1$.
That is a simple consequence of the symmetry
$J_b\to -J_b$ of the Gaussian distribution. The previous observations
entail in particular  (\ref{pippa}).
Consider in the torus $\Pi_\Lambda$ the interpolating potential
defined in (\ref{aip})
with the relative pressure $P^{(\Pi_\Lambda)}(t)$ and quenched state
$<->^{(\Pi_\Lambda)}_t$. Since
\be
\label{ifaq}
P^{(\Pi_\Lambda)}(0) \, = \, P^{(\Phi)}_{\Lambda} \; , \; \quad
P^{(\Pi_\Lambda)}(1) \, = \, P^{(\Pi)}_{\Lambda} \; ,
\ee
and
\be
P'(t) \, = \, \frac{1}{2}\sum_{b\in {\partial B(\Lambda)}}
(1-<q_b>^{(\Pi_\Lambda)}_t)  \; ,
\label{diff}
\ee
we have 
\be
P^{(\Pi)}_{\Lambda}-P^{(\Phi)}_{\Lambda} \, = \, \frac{1}{2}\sum_{b\in \partial
B(\Lambda)}\int_{0}^{1}(1-<q_b>^{(\Pi_\Lambda)}_t)dt \; ,
\label{diffmod}
\ee
which immediately entails theorem \ref{main2}.
\\\\
{\bf Proof of Theorem \ref{ht}}.  The cluster expansion
of \cite{Be} (see also \cite{FI,DKP}) overcomes the well known difficulty due 
the infinite
range of the Gaussian variable. We apply it to the present case to show that, 
regardless of the boundary conditions, each
$<q_b>$ is small for small
$\beta$ and definitely away from $1$.
Applying Proposition 1 of \cite{Be} to our problem (see in particular the proof 
of Lemma 3) we may
write 
\be
<q_b>^{(\Pi_{k\Lambda})}_t \; = \; A_{k\Lambda}(b,\beta^2,t)\b^2 +
C_{k\Lambda}(b,\beta^2,t) \; ,
\ee
where:
\newline
 (1) for every $\varepsilon$ we may choose
\be
|C_{k\Lambda}(b,\beta^2,t)| \; \le \; \frac{\varepsilon}{2} \; ,
\label{picc}
\ee
uniformly in all the variables and 
\newline
(2) $A_{k\Lambda}(b,\beta^2,t)$ is bounded uniformly in $(\Lambda,t)$ and is 
analytic in 
$\beta$ for 
$\beta <\b_0$, where
$\beta_0$ depends only on the dimension $d$ and not on $\Lambda$ and $t$. 
Remark once again that the parity
of the Gaussian variables yields the parity in
$\beta$ of each thermodynamic function so that the odd powers of the cluster
expansion vanish.

After integrating in $t$ we take the $k\to \infty$ limit of the previous
relation 
(which exists by Theorem 1 of \cite{Be} if $\beta <\b_0$), 
and  sum over all bonds in $\partial B(\Lambda)$. We obtain:
\be
\label{3.66}
\tau^{(\Phi)}_{\partial\Lambda} = -  \frac{\beta^2}{4}
\left[1-(A_\Lambda(\beta^2)\beta^2+C_\Lambda(\beta^2))\right] \; ,
\ee
with 
\be
C_\Lambda(\beta^2) \; = \; \lim_{k\to\infty}\int_{0}^{1}dt\frac{1}{|\partial
B(\Lambda)|}\sum_{b\in\partial
B(\Lambda)}C_{k\Lambda}(b,\beta^2,t)
\; ,
\ee
and
\be
A_\Lambda(\beta^2) \; = \; \lim_{k\to\infty}\int_{0}^{1}dt\frac{1}{|\partial
B(\Lambda)|}\sum_{b\in\partial
B(\Lambda)}A_{k\Lambda}(b,\beta^2,t)
\; ,
\ee
We remind that the multiplicative $\beta^2$ factor in (\ref{3.66}) comes from 
the fact
that the potential (\ref{tempe}) has interactions coefficients $\beta J$
whose variance is $\beta^2$. 
From (\ref{picc}) we derive the bound $|C_\Lambda|\le \varepsilon/2$. On the
other hand since the correlation
is bounded by one,\\ $|<q_b>^{(\Pi_{k\Lambda})}_t|\le 1$, and
$A_{k\Lambda}(b,\beta^2,t)$ is bounded uniformly in $(\Lambda,t)$, there is
$K>0$ independent of $\Lambda$ such that 
\be
\label{3.69}
|A_\Lambda(\beta^2)|<K \; .
\ee
Hence there is a $\overline{\b}>0$ such that  the quantity
$|\b^2A_{\Lambda}(\beta^2)|<C_1<1-\varepsilon/2$ if $\b<\overline{\b}$, 
uniformly in $\Lambda$. Hence, by (\ref{3.66}) we get the existence of $C>0$
independent of $\Lambda$ such
\be
\frac{\tau^{(\Phi)}_{\partial\Lambda}}{\beta^2}\; < -C\;<0
\ee
This proves Assertion (1). 

To prove assertion (2), remark that, given $\varepsilon >0$,  by (\ref{3.69})  
we can  always choose $\b{(\varepsilon)}$ in such a way 
that 
\be
|A_\Lambda(\beta^2)\beta^2| \; \le \; \varepsilon/2
\ee
uniformly with respect to $\Lambda$ if $\b<\b{(\varepsilon)}$.  Hence by 
(\ref{3.66}) we can conclude
\be
\frac{\tau^{(\Phi)}_{\partial\Lambda}}{\beta^2} \; \leq \; -  
\frac{1}{4}(1-\varepsilon)
\ee
if $\b<\b{(\varepsilon)}$ . 
The proof of
(\ref{m2}) is completely analogous.
\\\\
{\bf Outlook}\\
Our results show that the surface pressure has the expected {\it surface
size}
in dimension $d$. A change in the size dependence at low temperatures is very
unlikely. In fact our integral representation would force the
quenched overlap
 moments $<q>$ to be identically equal to one,
a situation which is not generally expected in the mean field picture
\cite{MPV} nor in the droplet one \cite{FH}.
A further step along the present line would be
the understanding of the variance of the difference of the pressure
computed with two boundary conditions, for example  periodic and
antiperiodic. This would yield a {\it surface tension} like
contribution. Bounds on the size dependence of such a quantity already exist
(see ref [74] in \cite{NS}) and it would be interesting
to investigate if the interpolating method can be used to obtain the correct
size; we hope to return elsewhere on that point 
and also on the existence
of the thermodynamic limit for the quenched surface pressure especially in
view to obtain an analogous of the second Griffiths inequality.\\\\
{\bf Acknowledgments}.
We thank M.Aizenman, A.Berretti, A.Bovier, A.C.D.van Enter, C. Giardina,
F.Guerra, J.Imbrie, C.Newman,
E.Olivieri and E.Presutti for interesting discussions.

\end{document}